\input phyzzx

\def\ls#1{_{\lower1.5pt\hbox{$\scriptstyle #1$}}}

\frontpagetrue


\let\picnaturalsize=N
\def\picsize{1.0in}
\def\picfilename{scipp_tree.eps}

\let\nopictures=Y

\ifx\nopictures Y\else{\ifx\epsfloaded Y\else\input epsf \fi
\let\epsfloaded=Y
{\line{\hbox{\ifx\picnaturalsize N\epsfxsize \picsize\fi 
{\epsfbox{\picfilename}}}\hfill\vbox{


\hbox{SCIPP 96/07} 
\vskip1.2in
}
}}}\fi


\overfullrule 0pt

\pubtype{ T}     

\rightline{SCIPP 96/07} \vskip-8pt


\title{Truly Strong Coupling and Large Radius in String Theory}
\author{Michael Dine and Yuri Shirman}
%
\address{Santa Cruz Institute for Particle Physics,
University of California, Santa Cruz, CA 95064}
\vskip-8pt



\vskip.4in
\vbox{
\centerline{\bf Abstract}

\singlespace
String theory, if it describes nature, is probably strongly
coupled.  In light of recent developments in string duality,
this means that the ``real world'' should correspond to a region
of the classical moduli space which admits no weak coupling
description.  We exhibit, in the heterotic string, one such
region of the moduli space, in which the coupling,
$\lambda$, is large
and the ``compactification radius'' scales as $\lambda^{1/3}$.
We discuss some of the issues raised by the conjecture that
the true vacuum lies in such a region.  These include
the question of coupling constant unification, and more
generally the problem of what quantities one might hope
to calculate and compare with experiment in such a picture.

}

\parskip 0pt
\parindent 25pt
\overfullrule=0pt
\baselineskip=19pt
\tolerance 3500
\endpage

\singlespace

\bigskip

\chapter{Introduction}

\REF\joereview{For a review, up to date as of Nov. 1995,
see J. Polchinski, {\it Recent Results in String Duality},
hep-th/9511157.}
\REF\wittencosmo{E. Witten, Mod. Phys.
Lett. {\bf A10} (1995) 2153;
hep-th/9506101.}

There is now compelling evidence for a variety
of dualitites between different string theories\refmark{\joereview}.
Strongly coupled regions of one string theory are
typically equivalent to weakly coupled regions of another,
and are thus solvable.  While it may be somewhat
premature, it is natural to ask:  what implications might
these observations have for the real world.

\REF\dineseiberg{
M. Dine and N. Seiberg,
Phys. Lett. {\bf 162B}, 299 (1985),
and in {\it
Unified String Theories}, M. Green and D. Gross, Eds. (World Scientific,
1986).}
Perhaps the most interesting possibility along
these lines is a conjecture by Witten concerning
the cosmological constant problem\refmark{\wittencosmo}.   But
there are other problems generic to any string phenomenology
which one might also hope to address.   One of these is the
problem of strong coupling\refmark{\dineseiberg}.
String theory, if it describes nature, is almost certainly
strongly coupled, since at weak coupling, one has runaway
behavior for the dilaton.  Duality does not, by itself, help
with this problem.  For if a region of strong coupling is equivalent
to a region of weak coupling in some other theory, then
it will suffer from the same sort of instability.

Indeed, what this suggests is that we should determine the
regions in the string moduli space which are {\it truly strongly
coupled}, i.e. which do not admit any perturbative description
at all.

\REF\vadim{V.S. Kaplunovsky,
Phys. Rev. Lett. {\bf 55} (1036) 1985.}
\REF\couplings{M. Dine
and N. Seiberg, Phys. Rev. Lett. {\bf 55} (366) 1985.}
\REF\bd{T. Banks and M. Dine, Phys. Rev. {\bf D50}
(1994) 7454;hep-th/9406132}
A second question in string phenomenology is the size of
any internal space, or the compactification radius, $R$.
While in the minimal supersymmetric standard model, unification
of couplings occurs at a scale well below $M_p$,
it is often said that the compactification scale must be
comparable to the string scale\refmark{\vadim,\couplings}.
But the argument for this is based on requiring that the
string coupling should be {\it weak}.  If the coupling is indeed
strong, one might imagine that the radius could be
very large.  Previous works have ignored this
possibility\refmark{\bd},
for a variety of reasons.
Now that we know more of these dualities, it should be
possible to explore this issue with greater precision.

In light of these two issues, it is natural to ask:  what are the
regions of the string moduli space which are truly strongly
coupled?  One might imagine that the answer would be:
string coupling of order one and string radius of order one.
But, as we shall see, there is a larger region, in which,
for example in the heterotic description, the couplings
becomes large, and the radius becomes large with
a certain power of the coupling.  We will not attempt, here,
to completely map out the strongly coupled regime of the
moduli space.  This would be a quite involved problem,
and is not really possible in any case, given the current
state of our knowledge.  Rather, our goal will be to show
that a regime of strong coupling and large radius
exists.

What are the implications of this observation?  After all, if no weak
coupling methods are available,
one might despair of ever being able to
say anything.  If the theory is strongly coupled
it is not even clear what one means by $R$ to start
with.  In theories with $N=4$ supersymmetry, the radius is related to the
mass of a set of BPS states, and so has a well-defined meaning,
even at strong coupling.  Moreover, the possible non-perturbative
dynamics in such theories are highly restricted by supersymmetry.
There can be no potential, for example.

In theories with
$N=1$ supersymmetry, the situation is more complicated.  The
states at weak coupling
with masses of order $1/R$ are not stable, and it is not
clear that they will correspond to any particular states at
strong coupling.  It is plausible that there should be states
with masses well below $M_p$, but this is only a guess.
On a more positive note, as
explained in ref. \bd, if the four dimensional coupling
is not too large (in the sense that $8 \pi^2 /g^2$ is large),
then even $N=1$ supersymmetry permits one to make a number
of statements about the theory.
\pointbegin
Because of the $2 \pi$ periodicity of the axion, stringy
non-perturbative effects in the superpotential and the gauge
coupling function are small.  The light spectrum
is the same as at weak coupling, and the theory
is approximately supersymmetric, up to
small effects (such as gluino condensation) which can be
seen in the low energy theory.
\point
Many important phenomena {\it are} controlled by inherently
stringy effects, which should receive large corrections
at strong coupling.  These are effects which are controlled
by the Kahler potential, and include:  stabilization of the
dilaton and other moduli (if it occurs) and the sizes of the soft breaking
terms.

\noindent
In ref. \bd, it was assumed that he radius was of order the string
scale; this was in part because of the authors'
belief that with fixed four dimensional coupling,
large radius would correspond, by some sort of duality, to weak coupling
and runaway behavior.  The observations of the present work suggest that the
radius could be significantly larger, and the coupling still strong.
We will not be able to offer any real explanation of why the coupling
and radius take the values they do.  Rather, we can only
observe that this is compatible
with our current understanding of string theory.

In the next section, we review the duality relations
for the different theories of interest, and discuss
the criteria for strong coupling.  In the third section we turn
to the problem of finding regions of strong coupling.
We will focus principally on theories in four dimensions
with $N=4$
supersymmetry.  We will see that if one takes the
ten-dimensional gauge coupling and the radius of the
heterotic string large but such that the four dimensional
coupling is fixed and of order one, perturbation theory
is not applicable to any of the dual descriptions.
The concluding section contains some speculations.  We
will conjecture that similar results hold for $N=1$ theories.
This is not an easy question to settle, since,
as we have just noted, the meaning of the
``compactification radius" of the strongly coupled theory
is not clear.  But there are other issues one must face
as well.  Unification of couplings suggests that the gauge
couplings at the high scale are small, and our
discussion above suggests that this is an essential
ingredient in any successful supersymmetry
phenomenology.  But we will see that if the four dimensional
coupling is very weak, perturbation theory is valid in the
Type I description of the theory (though not the Type II
description).  So, as in ref. \bd, we must assume that string
perturbation theory is already not viable for values of the
gauge couplings of the sort observed in nature.  As argued
there, this is plausible, but, with
the present state of our knowledge, it is certainly a
strong assumption.  Similar issues arise with the size of the
compactification radius.  At very large radius, one might
guess that any potential (even at strong coupling) should
vanish, e.g. due to ten dimensional supersymmetry.  We do not
know how to argue this rigorously, but suspect that the problem
of understanding why the radius is large is similar to that
of understanding why the four dimensional coupling is small.
Indeed, it is tempting to conjecture that this is another consequence
of string-string duality.

\chapter{Review of the Duality Relations}

\REF\wittenusc{E. Witten, Nucl. Phys. {\bf B443}
(1995) 85; hep-th/9503124.}
Matching of the low energy effective actions
between different string
theories requires  coupling-dependent rescalings of the
metric, and thus a rescaling of lengths\refmark{\wittenusc}.
Simply considering
the form of the world-sheet string action,
$$T\int d^2
\sigma \partial_{\alpha} X^{\mu} \partial_{\alpha} X^{\nu}
g_{\mu \nu}\eqn\twodaction$$
one sees that an overall rescaling of the
metric is equivalent to a rescaling of the
string tension (with lengths such as the compactification
scale held fixed).  Stated in this way, the duality mapping
between the heterotic theory and the Type I theory takes
the form:
$$T=e^{-\phi^{\prime}}T^{\prime}~~~~~e^{\phi}=e^{-\phi^{\prime}}
\eqn\typeone.$$
The argument can easily be made directly for the
low energy effective action as well.  Including the factors
of the string tension, but ignoring constants of order one,
the heterotic action takes the form:
$$\int d^{10} x e^{-2 \phi}T^4 (R + T^{-1} F_{\mu \nu}^2+ \dots)
\eqn\heteroticaction$$
while on the type I side one has:
$$\int d^{10} x e^{-2 \phi^{\prime}}T^{\prime 4} (R + e^{\phi^{\prime}}
T^{\prime -1}
F_{\mu \nu}^2+ \dots).\eqn\typeoneaction$$
The mapping of eqn. \typeone\ again takes one theory into the other.

\REF\aspinwall{P.S. Aspinwall and D.R. Morrison,
{\it String Theory on $K_3$ Surfaces}, preprint DUK-TH-94-68,
hep-th/9404151.}
\REF\joe{J. Polchinski, unpublished.}
Similar rescalings work for the heterotic--Type II case.  (We will
consistently use unprimed variables for the heterotic string, singly
primed variables for the Type I string, and doubly primed variables
for the Type II string).  Writing the six dimensional effective
action for the heterotic theory compactified on $T_4$, we have:
$$\int d^6 x e^{-2 \phi} T^4 v(R+ T^{-1} F_{\mu \nu}^2 + \dots)
= \int d^6 x e^{-2 \psi} T^2(R+T^{-1} F_{\mu \nu}^2+\dots)
\eqn\heteroticsix$$
where we have explicitly included a factor of the four dimensional
volume, and $e^{2 \psi}= e^{2 \phi}/(vT^2)$.
The Type II side is somewhat more complicated, since we need
to compactify on $K_3$.  It is simplest to work in the limit where
$K_3$ can be described as a $Z_2$ orbifold.  In this limit, the
gauge bosons appearing in the effective action come from
different sectors, and their kinetic terms have non-trivial
dependence on the moduli.  However, for the sixteen gauge bosons
which arise from twisted sectors, there are no factors of the volume
or the moduli, and one has:
$$\int d^6 x ( e^{-2 \phi^{\prime \prime}} T^{\prime \prime 4}
 v^{\prime \prime}
R+ 
T^{\prime \prime} F_{\mu \nu}^2 + \dots)
= \int d^6 x( e^{-2 \psi^{\prime \prime}} T^{\prime
\prime 2}R+
T^{\prime \prime} F_{\mu \nu}^2 + \dots)\eqn\kthreeaction$$
and $e^{2 \psi \prime \prime }= e^{2 \phi\prime \prime}/(vT^{\prime
\prime 2})$.  Now the lagrangians map into one another if
$$e^{2 \psi}=e^{-2 \psi^{\prime \prime}}~~~~~T=e^{-2 \psi^{\prime \prime}}
T^{\prime \prime}\eqn\typetwo.$$

To see that this works in detail, it is necessary to understand
how the radii on the heterotic and Type II sides
map into each other.  The general problem has been
discussed by Aspinwall\refmark{\aspinwall}.
Polchinski\refmark{\joe} has given a very explicit
mapping the case of the $Z_2$ orbifold (a special
case of $K_3$).  Taking, on each side, the underlying
tori to be products of circles, and calling the radii
on the heterotic side, $R_1$, $R_2$, $R_3$ and
$R_4$, and those on the Type II side $r_1$, $r_2$,
$r_3$ and $r_4$, this mapping is
$$R_1^2 T = r_1 r_2 / r_3 r_4$$
$$R_2^2 T = r_1 r_3 / r_2 r_4$$
$$R_3^2 T = r_1 r_4 / r_3 r_2$$
$$R_4^2 T= r_1 r_2 r_3 r_4 T^{\prime
\prime2}.\eqn\toruskthree$$
The inverse transformation is
$$r_1^2T^{\prime \prime}= R_1 R_2 R_3 R_4 T^2$$
$$r_4^2T^{\prime \prime} = {R_3 R_4 \over R_1 R_2}$$
$$r_3^2T^{\prime \prime} = {R_2R_4 \over R_1 R_3}$$
$$r_2^2T^{\prime \prime} = {R_1 R_4 \over R_2 R_3}\eqn\kthreetorus$$

It is easy to check that the various terms in the
effective action now map correctly into one another.  For
example, consider the gauge boson kinetic terms.  On the
heterotic side, at a generic point in the moduli
space, there are sixteen gauge
bosons in the Cartan subalgebra of $E_8 \times E_8$.
Their kinetic terms appear in the lagrangian with
coefficients $Te^{-2 \psi}.$  On the Type II side,
the orbifold possesses sixteen fixed points.  One
gauge boson appears at each fixed point.  The kinetic
term is independent of the coupling (these are Ramond-Ramond
fields) and of the radii (since they sit at the fixed points),
and so their coefficients are $T^{\prime \prime}$,
and, by virtue of equation \typetwo, they map simply
into each other.  Consider, next, the gauge bosons
which arise from untwisted sectors on the Type II side.
Their kinetic terms can be determined by dimensional
reduction from ten dimensions.  Six gauge
bosons arise from the three index
antisymmetric tensor, $A_{\mu IJ}$, with two indices
in the internal space.  If, say, $I=1,J=2$, the kinetic
term is proportional to $v^{\prime \prime}$, the
volume on the Type II side, and $g^{11}g^{22}$, or
$T^{\prime \prime}r_3 r_4 /r_1 r_2=R_{1}^{-2}
T^{\prime \prime}/T$.  This is the form of the
kinetic term for the gauge field arising from the
antisymmetric tensor $B_{\mu~4}$ on the heterotic
side.  Proceeding in this way, one can identify the mapping
of the other seven gauge bosons.  The factors of $T$
and $T^{\prime \prime}$ are crucial in getting all
of this to work.

In order to determine the regions of strong coupling,
it is necessary to understand what is the perturbative
expansion parameter in each theory.  We first proceed
in a very simple-minded way.
Suppose that all but $d$ dimensions
have been compactified on tori of radius $R$.
In the $d$ dimensional theory, loop amplitudes will involve
$$g_d^2 \sum_n\eqn\summomentum$$
where the sum is over momentum modes.  Changing the sum to
an integral, one obtains, approximately,
$$g_d^2 R^{10-d} \int d^{10-d}p.\eqn\sumintegral$$
The first factor is just the ten dimensional coupling
constant, in units of the string tension.

If some of the radii are small, the criterion is different
and can be determined from $T$-duality.  Suppose
$R_1 \dots R_a$ are small.  Perform a duality transformation
($R \rightarrow {1\over R}$) on each of these.  The $d$ dimensional
coupling constant is unchanged by this transformation, so the
ten-dimensional coupling must transform as
$$\tilde g^2 = g^2 \prod_{\rm small ~\rm radii} R_a^{-2}.\eqn\dualcoupling$$
So validity of perturbation theory requires that
$$g^2 \prod_{\rm small
~\rm radii} (R_a^2 T)^{-1}\eqn\smallradiuscondition$$
should be small.

\REF\pw{J. Polchinski and E. Witten,
{\it Evidence for Heterotic-Type I Duality}, IAS
preprint IASSNS-HEP95-81,hep-th/9510169.}
This criterion is correct for the Type II and heterotic theories,
but there are subtleties in the Type I case\refmark{\pw}.
The problem arises in the case that one dimension is much
smaller than the others.  Then
there are tadpoles for odd integer winding
states.  These leads to an extra factor of $1/(R^2 T^{\prime})$
in each order.  This factor is crucial in avoiding paradoxes
in dualities\refmark{\pw}.  If several dimensions are small
and of comparable size, this effect is not important\refmark{\pw}.

\chapter{Strong Coupling} 

The simplest approach to the problem of finding regions
of strong coupling is to
start with the heterotic string, and suppose that six dimensions
are compact, and of comparable size.   We will denote
these by $R$.  We will denote the heterotic string coupling
by $e^{\phi}$, and the string tension by $T$.
We will suppose that $R$ scales with the coupling as
$$R = e^{\alpha \phi}.$$
(Here and in what follows, if not indicated otherwise,
we will work in units of the heterotic string tension,
$T$.)  If $R^2 T>1$ (corresponding
to $\alpha > 0$), then strong coupling
certainly requires $e^{\phi}\gg 1$.    The case $R^2 T <1$
can be dealt with using $R \rightarrow 1/R$-duality.

First consider the implications of $S$ duality.
Under $S$ duality, the coupling, $e^{2 \phi}$,
transforms into
$$e^{2 \tilde \phi} = e^{-2 \phi} R^{12},\eqn\sduality$$
and so is strong provided $\alpha>1/6$.


The first question to ask is whether the radius on the Type
I side is large or small.   From eqn. \typeone, 
$$R^2 T^{\prime}= R^2 e^{-\phi}T,\eqn\areradiismall$$
so
$\alpha > 1/2$ corresponds to large radius, and the
coupling is necessarily weak.   If $\alpha < 1/2$, the
radius is small, and the condition for the validity of
perturbation theory is now that
$${e^{2 \phi^{\prime}} \over R^{12}} \ll 1.\eqn\typeonecondition$$
Using the relation between $T$ and $T^{\prime}$,
this says that
$\alpha \le {1\over 3}$.

Let us turn now to the duality to the Type II theory.  To make
the discussion simple, we will take the $K_3$ theory at the
orbifold point.  Note, first, from eqn.\kthreetorus,
we have
$$r_1^2= e^{4 \alpha \phi} (T^{\prime \prime})^{-1}
\eqn\kthreevolume$$
and
$$r_i^2 =(T^{\prime \prime})^{-1} ~~~~~i=1,2,3.\eqn\otherradii$$
Thus $r_1^2 T^{\prime \prime} \gg 1$, while
$r_i^2 \approx 1.$  We can determine the
coupling on the Type II side by using the relations:
$e^{2 \psi^{\prime \prime}}= e^{-2 \psi}$, so
$$e^{2 \phi^{\prime \prime}}/v^{\prime \prime}T^{\prime \prime 2}
= e^{-2 \phi} R^4 T^2 \eqn\determinesphi$$
giving
$$e^{2 \phi^{\prime \prime}}= e^{(-2+6 \alpha) \phi}.\eqn\typetwostrong$$
This is strong if $\alpha \ge1/3.$
So the theory is strongly coupled on all three sides
if (and only if) $\alpha = 1/3$.

$\alpha=1/3$ is a particularly interesting case.
On the heterotic side, it corresponds to taking the radius
and coupling constant large, while holding the four
dimensional coupling fixed and of order one.
What we have just learned is that in this limit, the expansion
parameters of both the Type I and Type II theories are of
order one.  In practice, as we have remarked in the
introduction, one would like the four dimensional coupling
to be somewhat small.  Being slightly more careful,
one finds that in this case the Type II expansion parameter
is of order $g_4^{-4/3}$, but the Type I expansion
parameter is of order $g_4^4$.  So if string
theory describes the real world, we must suppose that
the Type I perturbation theory is already not valid for
couplings which, from our field theory experience, seem
rather small.  Some arguments for this possibility were
advanced in ref. \bd.

Finally, one might ask if there is anything further which
can be learned from the eleven dimensional theory.\foot{We
thank E. Witten for raising this issue.}
In particular,
since the heterotic radius and coupling are large,
one might imagine that this corresponds to flat eleven dimensions.
However, if $\ell_{11}$
is the eleven dimensional Planck length, then
$$e^{2 \phi} = ({R_{11}\over\ell_{11}})^3 \gg 1 ~~~~~
R^2 T = e^{2 \phi/3} = {R^2 R_{11} \over \ell_{11}^3}.\eqn\elevendimensions$$
So, while the radius of the eleventh dimension is large,
the other ten dimensions satisfy
$${R^2 \over \ell_{11}^2} = e^{2 \phi/3}{\ell_{11} \over R_{11}}
\sim 1\eqn\elevenorderone$$
i.e. in Planck units, the radii are of order one.  Note that
if the four dimensional coupling is small, in the sense
described above, then the $11$ dimensional radius is large
(just as the coupling is weak in this case in the Type I theory).

One disturbing feature of this analysis is that there
do seem to be regions of weak coupling in the Type II
theory which are mapped to weak coupling in Type I.
As an example, consider compactifications
to six dimensions, where we compactify the
Type II theory on the (orbifold) $K_3$.
Suppose that the type II radii scale as
$$r_1= e^{\alpha \phi^{\prime \prime}}T^{\prime \prime~ -1/2} 
~~~~~r_i=e^{\beta \phi^{\prime \prime}}T^{\prime \prime ~-1/2},
i=2,3,4.
\eqn\weakweak$$
Then if $\phi^{\prime \prime}<0$ (weak coupling), $\alpha<0$
and $\beta>0$ (corresponding to three small and one large
radius on the Type II side) one finds that in the Type I description,
all of the radii are small, and that $R_4$ is much smaller than
the other three.  Using the criterion of ref. \pw, the
theory
is nominally weakly coupled if $\vert \alpha \vert > 2-3\beta$.
We suspect that in this case,
the condition for the validity of perturbation theory is
even stronger.  This question is currently
under study.

\chapter{Conclusions}

\REF\dualpairs{J.A. Harvey, D.A. Lowe
and A. Strominger, Phys. Lett. {\bf B362} (1995)
362, hep-th/9507168; C. Vafa and E. Witten,
{\it Dual String Pairs with N=1 and N=2
Supersymmetry in Four Dimensions} Harvard
preprint HUTP-95-A023, hep-th/9507050.}
We have seen that at least in theories with $N=4$
supersymmetry, one can take the coupling large
and scale $R \sim g^{\alpha}$ in such a way that the theory
is truly strongly coupled.  We would like to ask whether
such results apply to $N=1$ theories.  Niavely, one
might expect that they would, for example, in dual
pairs obtained by orbifolds of higher $N$
theories\refmark{\dualpairs}.  On the other hand,
while naively the scaling relations we have used
above should hold, it is not clear what they mean
in this case.   In particular, the momentum (or winding)
states are not BPS states in these theories, and their
masses would be expected to receive large corrections.
Still, we believe that the observations of the previous
suggestions make it plausible that the fundamental string 
coupling can be large, while the lowest string
thresholds can be below the Planck scale and the four
dimensional coupling can be of order one.

Let us suppose that these statements are true.
Then we can ask whether this region of the moduli
space could correspond to the true vacuum observed
in nature (assuming string theory {\it does} describe
nature).  While this region would not be amenable
to perturbative treatment, one might still worry
that one could establish that the potential for the
dilaton or other moduli tended to zero as the
coupling and radius tended to infinity, so that there
was still runaway behavior.  For example, if the theory
in this limit became flat, eleven dimensional space,
then eleven dimensional supersymmetry would be enough
to insure that there was not potential for the moduli
in this limit.  Hence there would be runaway behavior.
However, we noted in the previous section
that from the eleven dimensional perspective,
ten of the radii are ``Planckian'' in this limit.
Of course,
in the limit considered here, the theory looks like a flat
ten dimensional theory.  The leading terms in this theory
are determined by (ten-dimensional) supersymmetry, and
do not include a potential.  So any ten-dimensional potential,
$V_{10}(R)$
must vanish as some power of the radius.   On the other hand,
in four dimensional Planck units,
the four dimensional potential is given, naively, by
$$V_4=G_N^2 R^6 V_{10}.\eqn\fourdpotential$$
So this can be non-vanishing if $V_{10}$ does
not vanish more rapidly than $R^6$.  We do not see
at present how to establish such a strong bound on
the potential in the strong coupling limit, so we do not
believe one can rule out the possibility that the minimum lies
at some large value of the radius.  On the other hand, at weak
coupling, gaugino condensation gives a potential which falls
as $1/R^3$ for fixed $g_4$, and we suspect that one can
argue for a similar falloff at strong coupling.  If this
is the case, it is necessary to suppose that this falloff
sets in only for $R \gg M_p^{10}$, the ten dimensional Planck mass.
As weak support for this, it should be noted that
$RM_p^{10} \sim e^{\phi/12}$ in this limit, i.e. it grows
only very slowly with $\phi$.  This discussion also has
a flavor similar to that of ref. \bd, where it was argued
that string perturbation theory might break down for
smaller values of the coupling then expected from field
theory (we invoked this earlier to argue
that string perturbation theory might not be
valid in the Type I description, where the expansion parameter
is $g_4^2$).  Indeed, one might imagine that these two situations
are related by a string-string duality exchanging $S$ and $R$ ($T$).

If we suppose that these statements are true, then one can see the
outlines of a string phenomenology.  The superpotential
and gauge coupling functions must be analytic functions
of
$S={8 \pi^2/g_4^2}$. Moreover, at high scales they are
expected to periodic functions of $S$ with period $2 \pi$\refmark{\bd}. 
As a result, these functions are given exactly
by their one loop values, up to exponentially small
corrections.  Moreover, the spectrum is the same
as observed at weak coupling.  Thus one can imagine
starting with some four dimensional string model
at weak coupling, and reliably extracting a set
of predictions for the spectrum, some ratios of
Yukawa couplings, and coupling unification.
Any quantity which depended on the detailed values
of the Kahler potential could not be calculated in such
a scenario.  Such quantities would presumably include
the location of the minimum of the potential, the
soft breaking masses (at least their contributions
from high scale physics) and the cosmological
constant.  These would await methods for
treating the problem of ``truly strong coupling.''

We close with one possibly amusing note.  Many
authors have speculated, for various reasons, that
some internal radii might be very large,
while the others are small.  If one repeats the analysis
of section 3, with $D$ radii scaling as $e^{\alpha \phi}$,
while the four dimensional coupling is held fixed, one
finds that this situation is {\it always} mapped
to weak coupling if $D<6$.

\bigskip
\noindent
{\bf Acknowledgements.}
We thank Joe Polchinski for describing to us the map
between the Type II and heterotic theories, and for discussions
of many issues in duality.
We also thank Tom Banks, Nathan Seiberg, Steve Shenker
and Ed Witten for patient explanations
of many of the issues involved here.  M.D. is grateful
to the New High Energy Theory Center of Rutgers University
for its hospitality during the completion of this work.

\refout

\bye